\documentclass[preprint,12pt]{elsarticle}

\usepackage{amssymb}
\usepackage{color}

\journal{Physics of the Dark Universe}

\begin{document}

\begin{frontmatter}



\title{A tight correlation between the enclosed gravitational mass and hot gas mass in galaxy clusters at intermediate radii}


\author{Man Ho Chan}

\address{Department of Science and Environmental Studies, The Education University of Hong Kong \\ 
Tai Po, New Territories, Hong Kong, China}

\ead{chanmh@eduhk.hk}

\begin{abstract}
Many studies point out that there exists some tight correlations between dark matter and baryonic matter at different radii in galaxies. However, similar tight correlations can only be found in galaxy clusters for large radii. Here we report extremely tight correlations between the gravitational mass $M_{\rm grav}$ and hot gas mass $M_{\rm gas}$ in galaxy clusters at the hot gas core radius $r_c$ and at $2r_c$ (i.e. $M_{\rm grav}(r_c)$ vs $M_{\rm gas}(r_c)$ and $M_{\rm grav}(2r_c)$ vs $M_{\rm gas}(2r_c)$). By using the X-ray data of 64 large galaxy clusters with different sizes and masses, we find that the correlations can be described by a single relation $\log(M_{\rm grav}/M_{\odot})=(0.74 \pm 0.02) \log(M_{\rm gas}/M_{\odot})+(4.47 \pm 0.23)$ for a wide range of hot gas mass ($10^{11}M_{\odot}-10^{14}M_{\odot}$). The corresponding correlation coefficient and scatter are 0.97 and 0.10 dex respectively. This would be the first tight correlation with very small scatter between the enclosed gravitational mass and hot gas mass for galaxy clusters within intermediate radii ($\sim 100-1000$ kpc) and it provides a new kind of observational evidence to support the universality of correlation between dark matter and baryons. 
\end{abstract}


\end{frontmatter}



\section{Introduction}
Many studies reveal some close connections between dark matter mass (or gravitational mass) and baryonic mass in galaxies. For example, early observations discovered a tight correlation between the luminosity (determined by baryonic mass) and the stellar rotational velocity (affected by gravitational mass). It is known as the Tully-Fisher relation \cite{Tully,McGaugh,Lelli2}. This relation challenges the current models of galaxy formation and it does not agree with the prediction from the cold dark matter (CDM) cosmology \cite{Lelli2}. Besides, some studies find that the combination of the contributions of dark matter and baryons results in an almost flat rotation curve in many galaxies. It seems that dark matter `cooperates' with baryonic matter to achieve a flat galactic rotation curve \cite{Battaner,Remus}. Furthermore, the mass-discrepancy-acceleration (MDA) relation \cite{McGaugh2}, the central-surface-densities relation \cite{Lelli} and the radial acceleration relation (RAR) \cite{McGaugh3,Li} in galaxies seem to indicate some tight correlations between dark matter mass and baryonic mass. In particular, the RAR reveals a striking correlation between the acceleration observed (determined by gravitational mass) and the distribution of baryons at every radius for 175 galaxies \cite{McGaugh3}. The resulting scatters are very small ($\approx 0.13$ dex) and largely dominated by observational uncertainties. The RAR is a very important discovery because it indicates a very general tight correlation between gravitational mass and baryonic mass at different small and intermediate radii, not just for large radii of galaxies. Although some studies can produce these relations based on the CDM model \cite{Desmond,Chan,Ludlow}, some of the parameters involved have to be adjusted to match the observed properties of galaxies. Therefore, a satisfactory explanation is still required to account for the observed strong coupling between dark matter and baryons. 

Interestingly, these tight relations with small scatters have not been not found explicitly at intermediate radii for galaxy clusters. In fact, many relations have been discovered for galaxy clusters. For example, numerical simulations have shown how hot gas fractions in galaxy clusters depend on their redshifts, radii and total mass of galaxy clusters \cite{Springel,Planelles}. Observations have also obtained these relations \cite{Lin,Giodini2,Lagana2,Planelles}. In particular, some correlations between total gravitational mass and total hot gas mass of galaxy clusters have been obtained \cite{Sanders,Vikhlinin2,Mantz,Mahdavi,Donahue,Mantz2}. However, most of the correlations obtained are considering the total gravitational mass at $r_{500} \sim 1 $ Mpc ($r_{500}$ is the radius where the average mass density equals 500 times of the cosmological critical density), but not the enclosed gravitational mass at intermediate radii. Although some recent studies have examined the details of galaxy clusters at intermediate radii (e.g. at $r_{2500}$) \cite{Schellenberger,Eckert}, the explicit correlations between dark matter and baryonic matter at intermediate radii have not been widely discussed to understand their potential coupling. Besides, some scaling relations relating gravitational mass and properties of hot gas (e.g. luminosity, temperature or baryon fraction) in galaxy clusters have been obtained \cite{Ota,Chen,Sun,Ventimiglia,Lagana,Giodini,Ettori}. However, most of the resulting relations found quite depend on the types of galaxy clusters \cite{Chen} and the resulting scatters are pretty large.

In this article, we report a surprising discovery of a tight correlation between the enclosed gravitational mass and hot gas mass of galaxy clusters within 2 times of the hot gas core radius (intermediate radii $r \sim 100-1000$ kpc) for a wide range of hot gas mass. We use the X-ray data of 64 large galaxy clusters (core radius $\ge 100$ kpc) and calculate the enclosed gravitational mass and hot gas mass at core radius $r_c$ and 2 times of the core radius ($2r_c$) for each of the galaxy clusters. We find a tight single power-law correlation between them with a scatter as small as 0.10 dex and a correlation coefficient = 0.97. In the followings, the cosmological model assumed is a flat $\Lambda$CDM model with Hubble constant $H_0=68$ km s$^{-1}$ Mpc$^{-1}$, $\Omega_m=0.27$ and $\Omega_{\Lambda}=0.73$.

\section{Method}
The X-ray surface brightness distribution can be obtained precisely by X-ray observations \cite{Reiprich,Chen}. The surface brightness profile can be described by a $\beta$-model:
\begin{equation}
S(R)=S_0 \left[1+ \left(\frac{R}{r_c} \right)^2 \right]^{-3\beta+1/2},
\end{equation}
where $S_0$ is the central surface brightness, $R$ is the projected radius, $r_c$ is the core radius and $\beta$ is a fitted parameter. This surface brightness profile can be transformed to a number density profile which describes the number density distribution of hot gas in a galaxy cluster:
\begin{equation}
n(r)=n_0 \left[1+ \left(\frac{r}{r_c} \right)^2 \right]^{-3\beta/2},
\end{equation}
where $n_0$ is the central number density. Assuming the hot gas is in hydrostatic equilibrium, we get
\begin{equation}
\frac{dP(r)}{dr}=-\frac{GM_{\rm grav}(r)m_gn(r)}{r^2},
\end{equation}
where $P(r)=n(r)kT$ is the hot gas pressure, $T$ is the hot gas temperature, $M_{\rm grav}(r)$ is the gravitational mass profile and $m_g$ is the average mass of a hot gas particle. By combining Eq.~(2) and Eq.~(3), we have
\begin{equation}
M_{\rm grav}(r)=-\frac{kTr}{G\mu m_p} \left(-3\beta \frac{r^2}{r^2+r_c^2}+ \frac{d \ln T}{d\ln r} \right),
\end{equation}
where $\mu=0.59$ is the molecular weight and $m_p$ is proton mass. Some studies have pointed out that the hydrostatic mass calculated may not be equal to the cluster gravitational mass \cite{Hoekstra}. Simulations have also found this hydrostatic bias at intermediate radii \cite{Biffi}. Although a recent study shows that the X-ray hydrostatic mass measurements are remarkably robust and method-independent \cite{Bartalucci}, the hydrostatic bias in galaxy clusters seems to be non-negligible. This will contribute some systematic uncertainty in the analysis ($\sim 15-20$\%) \cite{Biffi}. Furthermore, recent studies show that the hydrostatic bias is mass-dependent and the hydrostatic mass is a scattered proxy of the real cluster gravitational mass \cite{Schellenberger}. For $M_{\rm grav} \ge 5 \times 10^{14}M_{\odot}$, the systematic bias could be as large as 40\% \cite{Schellenberger}. However, as shown later, the objective of our study is not going to examine the accurate quantified relation between the hot gas mass and the gravitational mass, but the potential correlation between them. The systematic deviation between hydrostatic mass and the real gravitational mass might slightly affect the scatter in the correlation. Nevertheless, the systematic deviation will not reproduce a tight correlation if they are not indeed correlated.

We consider the gravitational mass at two different positions (at $r=r_c$ and $r=2r_c$) to examine the behavior of $M_{\rm grav}$ in different regions (central and outer regions of hot gas) respectively. Generally speaking, the core radius $r_c$ characterizes the transition of the radius dependence of the gravitational mass and hot gas density functions. It also defines the sizes of hot gas and dark matter halo of a galaxy cluster. Therefore, we use $r_c$ as a reference position. Moreover, it can be justified that $r=r_c$ and $r=2r_c$ are robust positions to do the analysis. First, there are several galaxy clusters which have $2r_c<r_{500}<3r_c$ \cite{Chen}. Therefore, choosing too large $r$ (e.g. $r \ge 3r_c$) would probably outweigh the effective radius of these galaxy clusters. Also, the systematic uncertainties of the surface brightness profile would be significant if we consider too large $r$. For small $r$, we choose $r=r_c$ because it characterizes the gravitational mass within a `cored region' of a galaxy cluster. Besides, the uncertainties of $T$ would be significant if $r$ is too small. For example, the uncertainties of $T$ for $r<r_c$ are quite large in the Perseus cluster \cite{Churazov}. Therefore, considering the gravitational mass and hot gas mass at $r=r_c$ and $r=2r_c$ can minimize the systematic uncertainties and almost represent the enclosed mass within a major part of a galaxy cluster. 

The temperature profile for a galaxy cluster is almost constant, except for the central regions of cool-core clusters \cite{Sanderson,Hudson}. The temperature change in the outer regions ($r_c \le r \le 0.5r_{500}$) is less than 15\% \cite{Vikhlinin,Reiprich2}. Therefore, we set $d\ln T/d \ln r=0$ for all non-cool-core clusters. The systematic uncertainty for this assumption is less than 15\% \cite{Hudson}. For cool-core clusters, although observations indicate that the slopes of their central temperature profiles are quite universal and close to a constant $d\ln T/d \ln r \approx 0.4$ \cite{Sanderson}, the temperature profiles for individual clusters may still have difference. Some studies point out that the central temperature of the cool-core clusters is approximately half of the bulk component temperature \cite{Ikebe,Chen}. However, the uncertainty is still large if we take this approximation. In the following analysis, we are going to consider the gravitational mass within 2 times of the core radius ($r \le 2r_c$). Since the rise of hot gas temperature in a cool-core cluster mainly occurs in the central region, we take $d \ln T/d \ln r=0$ at $r=2r_c$ for the cool-core clusters and we neglect the data for the 12 cool-core clusters at $r=r_c$. In other words, for our 64 sample clusters (see below), we consider the gravitational mass of 52 non-cool-core clusters at both $r=r_c$ and $r=2r_c$, and 12 cool-core clusters at $r=2r_c$ only. The temperature profiles are all taken as constant $T=T_h$, where $T_h$ is the temperature of the bulk component \cite{Chen}. As a result, the gravitational mass $M_{\rm grav}(r_c)$ and $M_{\rm grav}(2r_c)$ depends on the values of $\beta$, $T$ and $r_c$ only.

On the other hand, we can calculate the hot gas mass profile by integrating Eq.~(2):
\begin{equation}
M_{\rm gas}(r)=m_gn_0 \int_0^r 4\pi r'^2 \left(1+ \frac{r'^2}{r_c^2} \right)^{-3 \beta/2}dr'.
\end{equation}
The above equation can be simplified to $M_{\rm gas}(r)=4\pi m_gn_0r_c^3I(\beta,x)$, where $I(\beta,x)=\int_0^xx'^2(1+x'^2)^{-3\beta/2}dx'$ with $x=r/r_c$. The integral $I(\beta,x)$ for $\beta=0.4-1.0$ at $r=r_c$ ($x=1$) and $r=2r_c$ ($x=2$) can be approximately given by $I(\beta,1)=0.1764 \beta^{-0.4119}$ and $\log I(\beta,2)=-0.9348(\log \beta)^2-1.352 \log(\beta)-0.2624$. Therefore, the baryonic mass $M_{\rm gas}(r_c)$ and $M_{\rm gas}(2r_c)$ depends on the values of $n_0$, $\beta$ and $r_c$ only.

\section{Result}
We use a sample of low-redshift ($z=0.0037-0.2$) galaxy clusters (the extended HIFLUGCS, altogether 106 galaxy clusters) in \cite{Chen} to perform the analysis. To minimize the systematic uncertainties, we only choose 64 large galaxy clusters ($r_c \ge 100$ kpc). These clusters include cool-core and non-cool-core clusters with a wide range of masses ($\sim 2\times 10^{13}M_{\odot}-4 \times 10^{15}M_{\odot}$), temperature ($T \sim 1-10$ keV) and sizes ($r_{500}\sim 0.8-2.9$ Mpc). We neglect small galaxy clusters ($r_c<100$ kpc) because many of them are dominated by the bright cluster galaxies (BCGs), which may induce large systematic uncertainties in gravitational mass determination. One may argue that the X-ray flux limited samples in HIFLUGCS may preferentially include core-luminous clusters (the cool-core bias) \cite{Rossetti}, which has selection effects to affect the results.  This would also lead to the so-called Malmquist-Eddington bias which makes our results less representative \cite{Teerikorpi}. However, as mentioned above, only 12 cool-core clusters ($\sim 20$\%) are included in our sample, which is close to the fraction of cool-core clusters ($29 \pm 4$\%) investigated in the Planck sample \cite{Rossetti}. The observational data used in this study are based on ROSAT and ASCA observations, which may have poor angular resolution in the inner regions. This is also the reason why we consider large galaxy clusters ($r_c \ge 100$ kpc) and investigate the correlations in the intermediate radii. The effects of bias due to this problem would be much alleviated.

Among these 64 clusters, 27 hot gas profiles are better fitted with a double-$\beta$ model \cite{Chen}. Nevertheless, the reduced $\chi^2$ values of using the single-$\beta$ model and the double-$\beta$ model are close to each other for these 27 clusters. Only one cluster (A3558) has the reduced $\chi^2$ value greater than 3 for the single-$\beta$ model \cite{Chen}. Note that the data quality used in \cite{Chen} is not very high so that some systematic errors are introduced when the fits are performed over large radii. Although using the single-$\beta$ model may introduce some systematic errors, we have to use the single core radii $r_c$ as reference radii of our sample clusters for consistency (two core radii appeared in the double-$\beta$ model for each cluster). Furthermore, recent studies show that the flux in the outskirts is systematically under-predicted by using a single-$\beta$ model \cite{Kafer}. This problem is more severe for strong cool-core clusters. However, for non-cool-core or weak-cool-core clusters, the accuracy of the flux in the outskirts stays at the $\sim 4$\% level for these clusters \cite{Kafer}. In our 64 sample clusters, we have only 12 cool-core clusters and only 4 of them are strong cool-core clusters. Also, we only consider up to $r=2r_c$, which is not very outskirt region. Therefore, we still follow the single-$\beta$ model to perform the analysis and the systematic errors of using the single-$\beta$ model for the 27 clusters have to be considered.  

Putting the fitted parameters $n_0$, $\beta$, $r_c$ and $T$ with uncertainties in \cite{Chen} into Eq.~(4) and Eq.~(5), we can get the gravitational mass and baryonic mass at $r_c$ and $2r_c$ for each of the galaxy clusters. Note that the parameters fitted in \cite{Chen} have assumed the Hubble parameter $h=0.5$. We have re-scaled the parameters to match the current value $h=0.68$ \cite{Planck}. By plotting $\log M_{\rm grav}$ against $\log M_{\rm gas}$ at $r=r_c$ and $r=2r_c$ respectively, surprisingly they lie on the same nearly perfect straight line (see Fig.~1):
\begin{equation}
\log \left( \frac{M_{\rm grav}}{M_{\odot}} \right)=(0.74 \pm 0.02) \log \left(\frac{M_{\rm gas}}{M_{\odot}} \right)+(4.47 \pm 0.23).
\end{equation}
This tight correlation has not been discovered previously. It is the first tight correlation for a wide range of enclosed hot gas mass ($10^{11}M_{\odot}-10^{14}M_{\odot}$) that systematically describes the mass relation between dark matter and hot gas mass in a large sample of galaxy clusters at intermediate radii ($\sim 100-1000$ kpc). Previous correlations obtained are usually with large uncertainties and they are only valid for a small range of hot gas mass at large radii ($\sim 10^{13}M_{\odot}-10^{14}M_{\odot}$) \cite{Sanders,Vikhlinin2,Mantz,Mahdavi,Donahue,Mantz2}. Four galaxy clusters in our sample have relatively high redshift $z=0.1-0.2$ and we have found no special redshift trend in the correlation, though the size of this sub-sample is very small. Fig.~2 shows the distribution of the residuals and it is well described by a Gaussian function of width $\sigma=0.09$ dex. The resulting root-mean-square scatter and the correlation coefficient are 0.10 dex and 0.97 respectively, which indicates a very strong correlation between $M_{\rm grav}$ and $M_{\rm gas}$ for a wide range of gravitational mass. The total expected scatter is 0.14 dex (see Table 1), which means nearly no room for intrinsic scatter. Here, the expected scatter mainly comes from the uncertainties of measurement presented in \cite{Chen}. For instance, the uncertainties of $T$ originate from median X-ray luminosity observations and the uncertainties of $\beta$, $r_c$ and $n_0$ originate from the fitting of X-ray surface brightness profiles. Therefore, the uncertainties of the parameters do not have strong correlations and the error bars in Fig.~1 reflect these uncertainties. Besides the measurement errors, we also include the maximum possible systematic uncertainties of $d \ln T/d\ln r$ in the calculations of the expected scatter. Nevertheless, as mentioned above, the bias in our sample selection and the involved systematic uncertainties are not completely negligible so that the existence of the intrinsic scatter in the sample is still possible.

In most of the galaxy clusters, the intergalactic hot gas mass dominates the baryonic mass \cite{Cavaliere}. Generally speaking, the total intergalactic hot gas mass is about 5-10 times larger than the total stellar mass in galaxies inside a galaxy cluster \cite{Lagana}. Therefore, the spherical hot gas component in a galaxy cluster can almost represent the total baryonic mass. Apart from the possible unknown interplay between dark matter and baryons, the correlation can be explained by two possible natural explanations: 1. it is related to the cosmological baryon fraction ($f=M_{\rm gas}/M_{\rm grav} \sim 0.16$); 2. the correlation is caused by standard gravitation. In Fig.~3, we plot the baryon fraction $f$ against $M_{\rm gas}$ and we can see that this ratio is not a constant. Based on our fitted empirical relation, we get $f \propto M_{\rm gas}^{0.26}$. In particular, the cored regions (at $r_c$) of many galaxy clusters are much dominated by dark matter ($f \le 0.1$). This is expected because the effects of feedback might be easier to push gas out to larger radius. However, for outer radii, the values of $f$ is larger and approach to the cosmological baryon fraction. Therefore, the values of $f$ are varying in different positions and it is difficult to explain the entire correlation for a wide range of $M_{\rm gas}$ by relating it to a single value of the cosmological baryon fraction. 

For the second explanation, we try to reproduce the empirical correlation using the standard CDM model. Using the Navarro-Frenk-White (NFW) dark matter density profile predicted from CDM simulations \cite{Navarro} and following Eqs.~(3) and (5) (assuming constant $T \sim 3-8$ keV), we can determine $M_{\rm gas}$ and $M_{\rm grav}$ separately at different radii. By randomly choosing some values of $n_0$, scale radius $r_s=100-500$ kpc and constraining $f=0.16$ at 1 Mpc, we can get a relation between $M_{\rm grav}$ and $M_{\rm gas}$ at two different radii (100 kpc and 500 kpc) (see Fig.~1). The resultant power-law fitted is $M_{\rm grav} \propto M_{\rm gas}^{0.53}$, which is somewhat deviated from our empirical correlation. The root-mean-square scatter is 0.24 dex, which is much larger than the scatter of the empirical correlation. The actual scatter would be much larger if we release the constraint of $f$ at 1 Mpc (the actual range of $f$ at 1 Mpc can be as large as 0.05-0.2; see Fig.~3). We also model the dark matter density profile by the Einasto profile, which is believed to be compatible with the CDM model \cite{Javid}. Using similar parameters and constraints, we get a power law $M_{\rm grav} \propto M_{\rm gas}^{0.59}$ with a scatter 0.25 dex (see Fig.~1), which is close to that obtained by using the NFW profile. Therefore, by comparing the power laws and the scatters, the explanation of the obtained correlation using standard CDM model is not favoured.

\begin{table}
\caption{Scatter budget for the correlation.}
 \label{table1}
 \begin{tabular}{@{}lc}
  \hline
  Source &  Residual \\
  \hline
  Errors in $T$ & 0.08 dex\\
  Errors in $r_c$ & 0.08 dex \\
  Errors in $\beta$ & 0.06 dex \\
  Errors in $n_0$ & 0.06 dex \\
  \hline
  Total & 0.14 dex \\
  \hline
 \end{tabular}
\end{table}

\begin{figure}
\vskip 10mm
 \includegraphics[width=140mm]{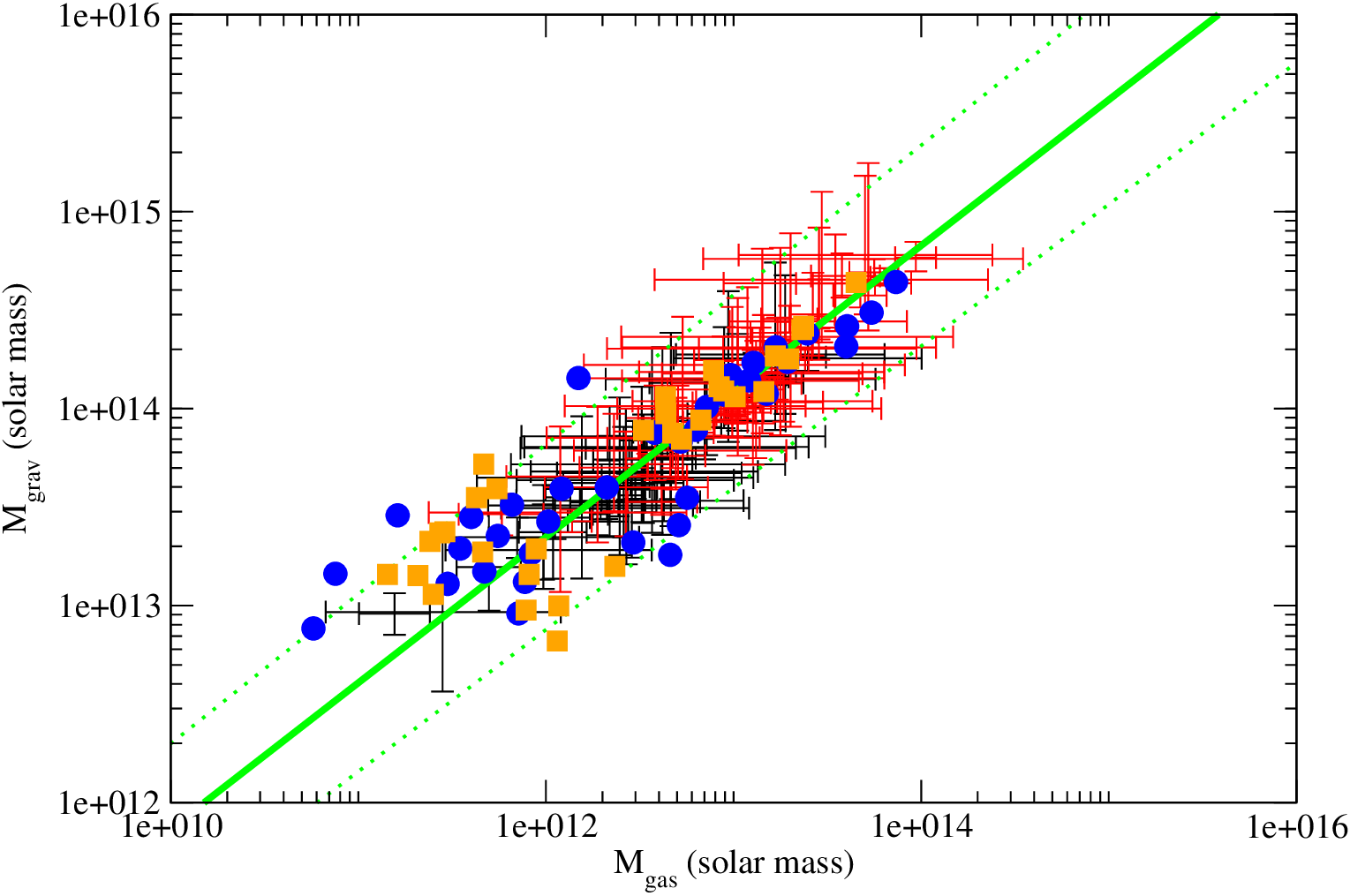}
 \caption{The data points (total 116) with error bars indicate the relations of $M_{\rm grav}$ and $M_{\rm gas}$ for 64 galaxy clusters (black: at $r=r_c$; red: at $r=2r_c$) \cite{Chen}. The green solid line is the best-fit straight line $\log (M_{\rm grav}/M_{\odot})=0.74 \log (M_{\rm gas}/M_{\odot})+4.47$. The two green dotted lines indicate the 1$\sigma$ limits for the best-fit line. The blue circles and orange squares represent the data predicted by the CDM model, using the NFW profile and the Einasto profile, respectively.}
\vskip 10mm
\end{figure}

\begin{figure}
\vskip 10mm
 \includegraphics[width=140mm]{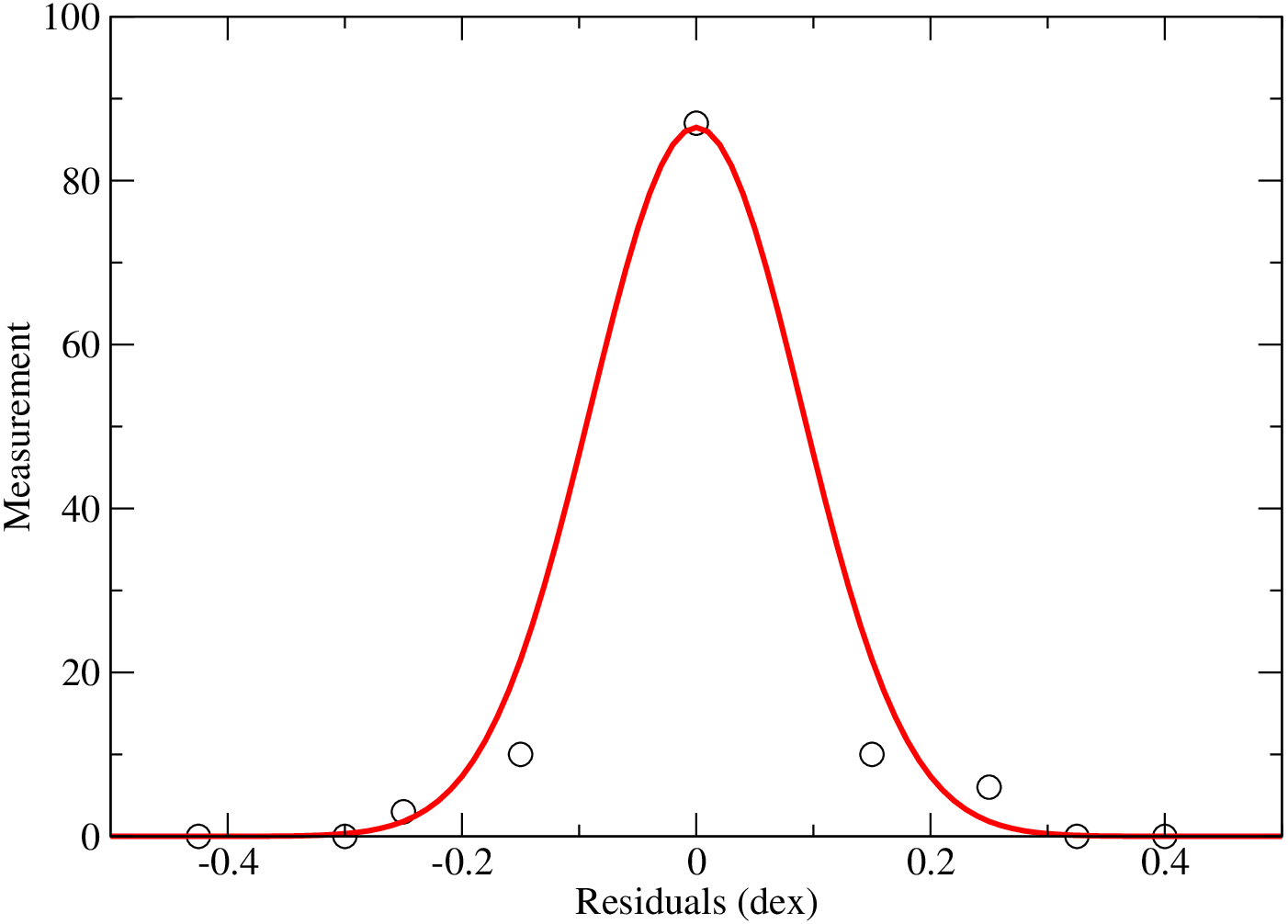}
 \caption{The distribution of the residuals of $\log M_{\rm grav}$ (indicated by circles). The red line is the Gaussian fit of the distribution with $\sigma=0.09$ dex.}
\vskip 10mm
\end{figure}

\begin{figure}
\vskip 10mm
 \includegraphics[width=140mm]{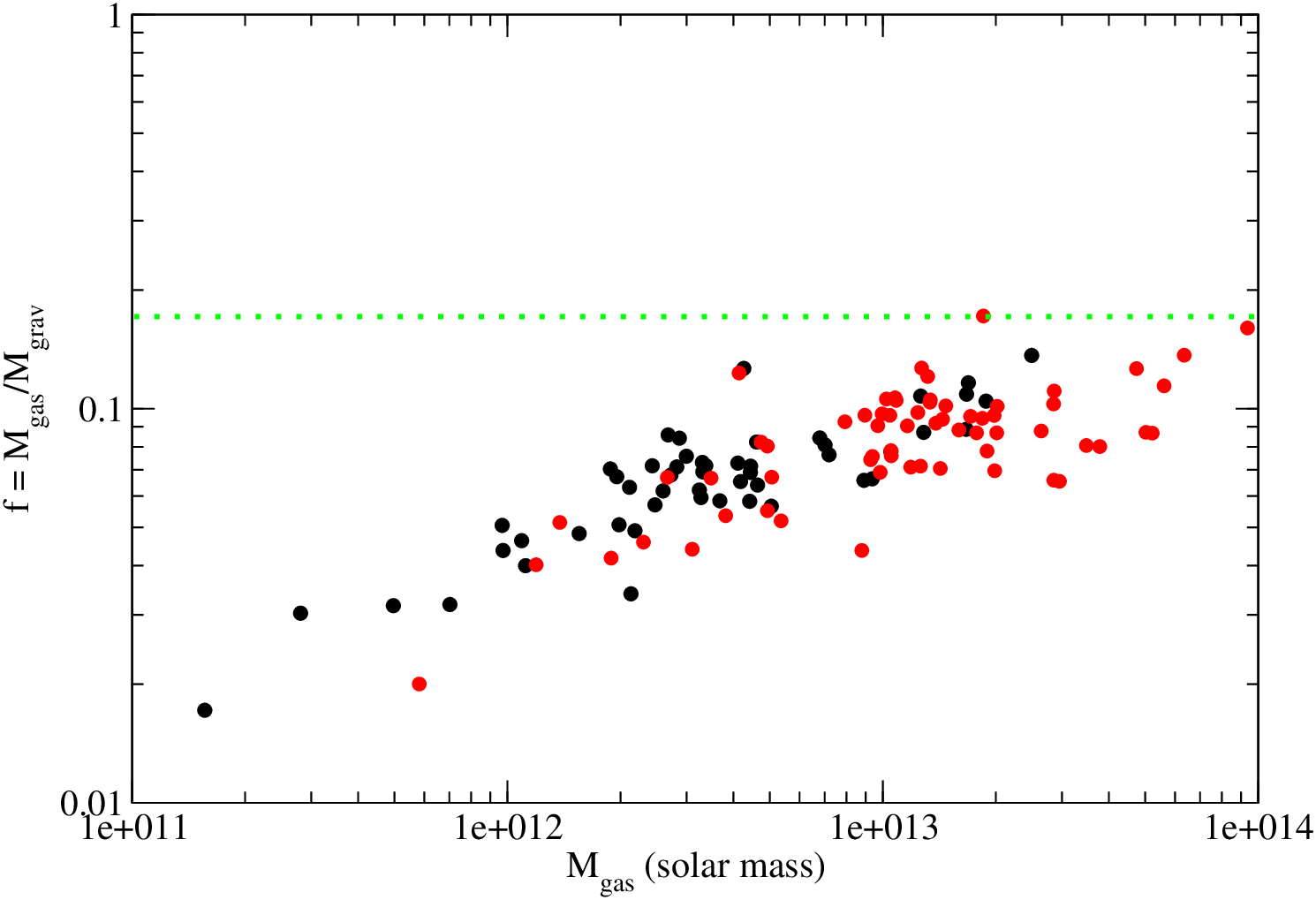}
 \caption{The graph of $f$ against $M_{\rm gas}$ (black: at $r=r_c$; red: at $r=2r_c$). The green dotted line represents the cosmological baryon fraction $f=0.16$.}
\vskip 10mm
\end{figure}

\begin{figure}
\vskip 10mm
 \includegraphics[width=140mm]{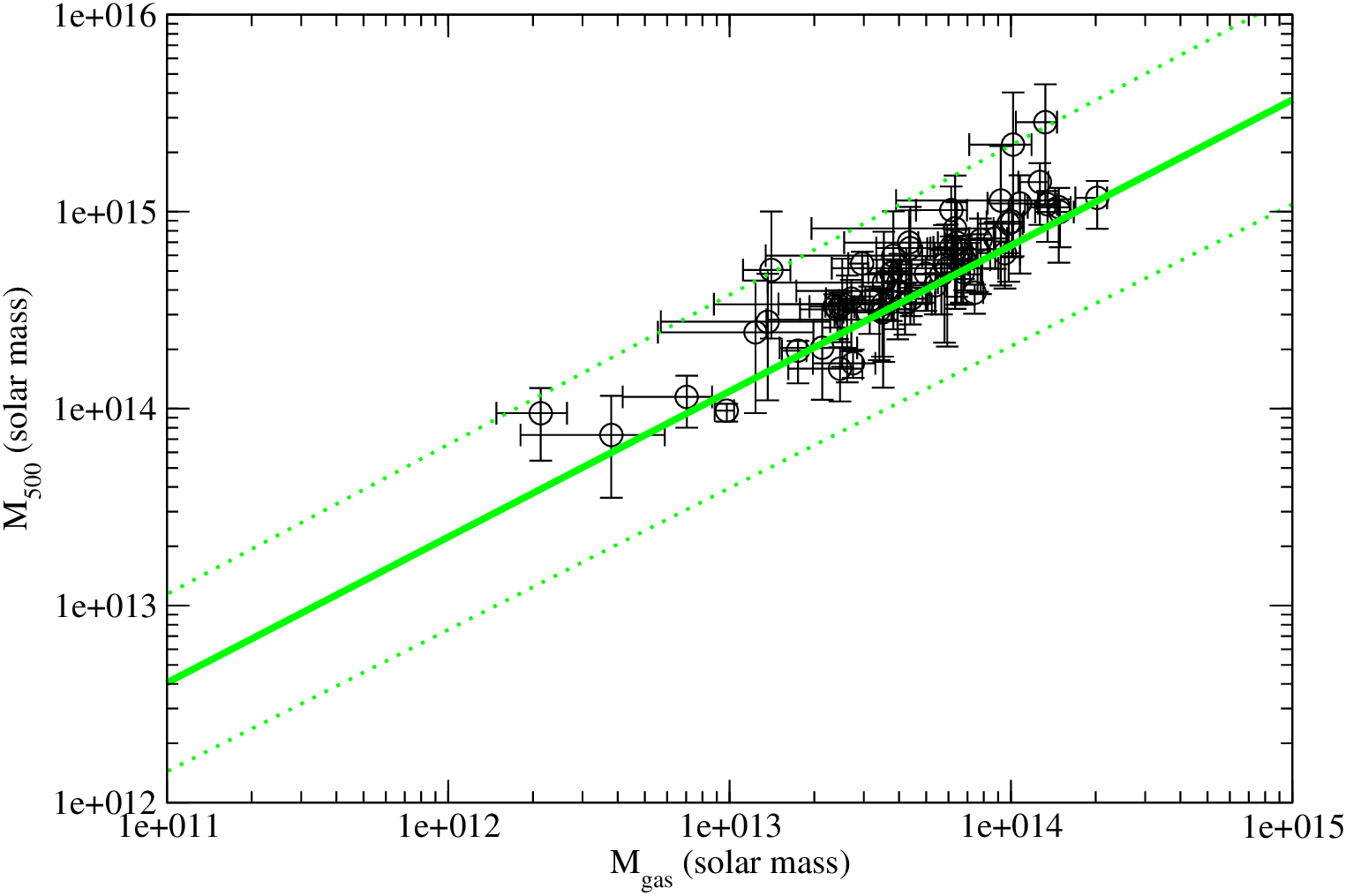}
 \caption{The black data points with error bars indicate the relations of $M_{500}$ and $M_{\rm gas}$ (at $r_{500}$) for the 64 sample galaxy clusters \cite{Chen}. The green solid line is the best-fit straight line for the correlation at intermediate radii $\log (M_{\rm grav}/M_{\odot})=0.74 \log (M_{\rm gas}/M_{\odot})+4.47$. The two green dotted lines indicate the 1$\sigma$ limits for the best-fit line.}
\vskip 10mm
\end{figure}

\begin{figure}
\vskip 10mm
 \includegraphics[width=140mm]{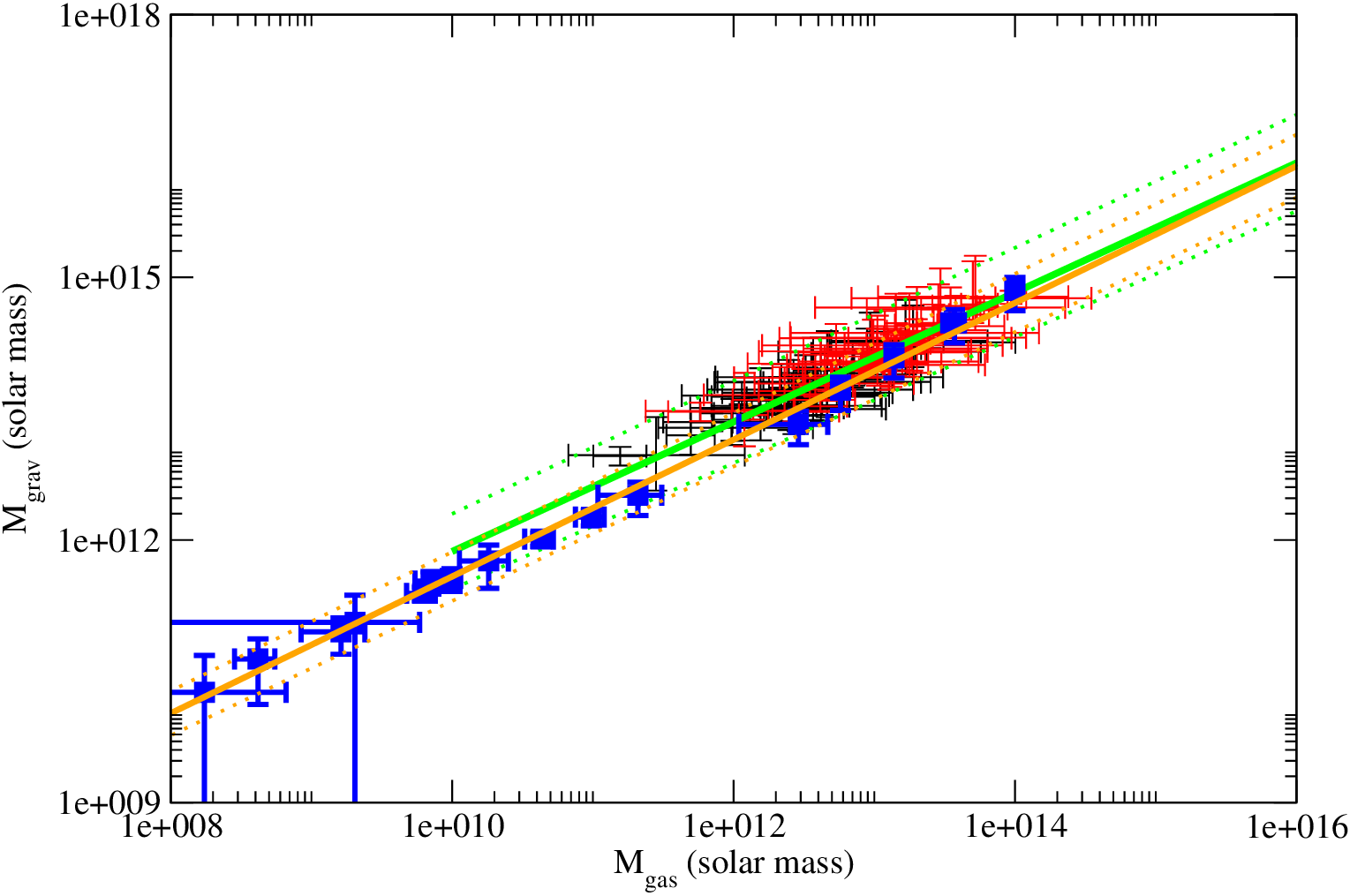}
 \caption{A similar plot of Fig.~1 for the correlation. The blue squares and the orange solid line represent the correlation between baryonic and gravitational masses of different cosmic structures \cite{McGaugh4}. The orange dotted lines indicate the 1$\sigma$ limits for the correlation. Here, we assume that $M_{\rm gas}$ represents the baryonic mass $M_{\rm bar}$.}
\vskip 10mm
\end{figure}

\begin{figure}
\vskip 10mm
 \includegraphics[width=140mm]{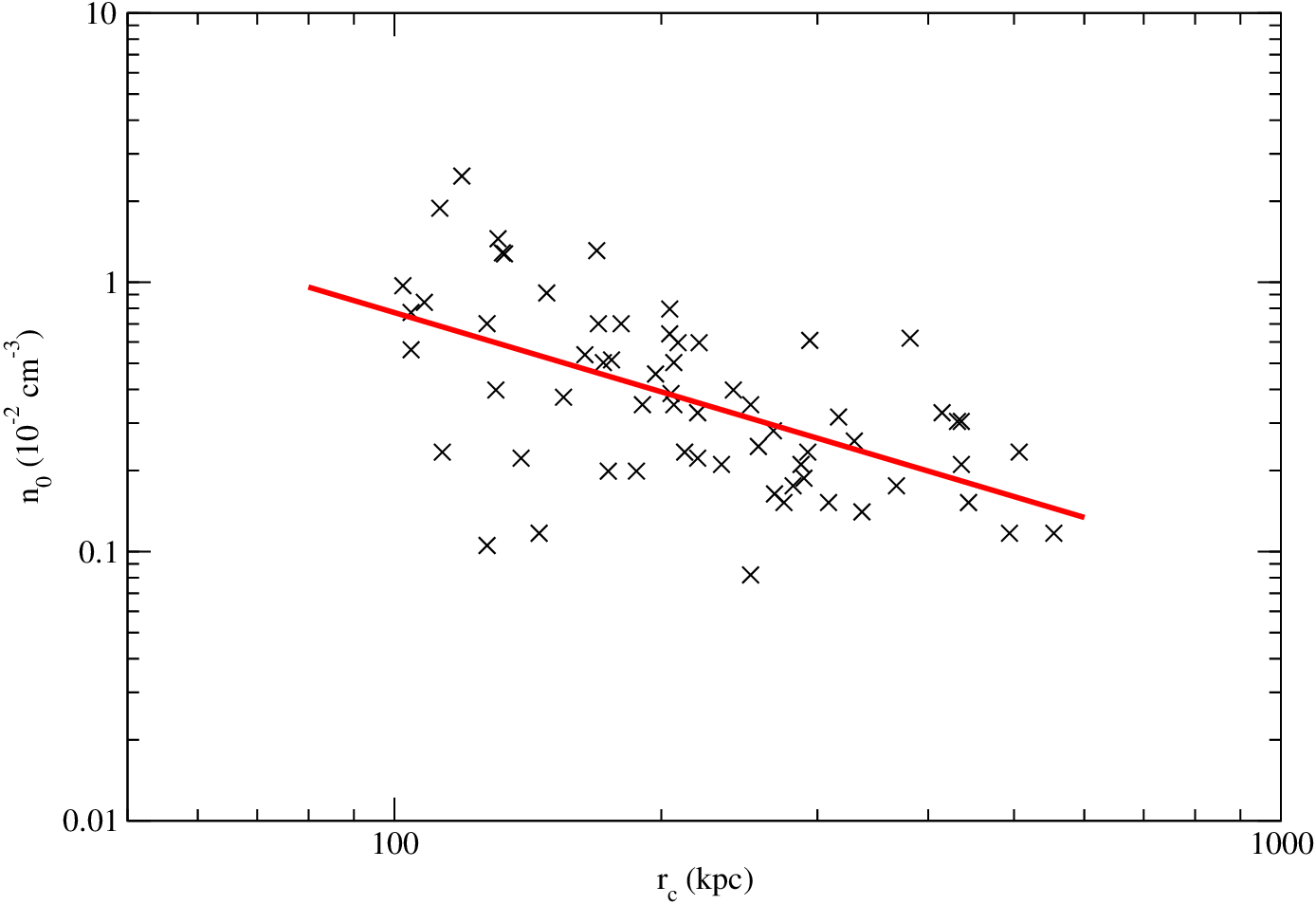}
 \caption{The correlation between $n_0$ and $r_c$. The scatter of the correlation is 0.27 dex.}
\vskip 10mm
\end{figure}

As mentioned above, many studies have examined the correlation between the total gravitational mass ($M_{500}=M_{\rm grav}(r_{500})$) and total hot gas mass \cite{Sanders,Vikhlinin2,Mantz,Donahue,Mantz2}. Some studies also discuss the relation between baryon fraction and total gravitational mass, which are equivalent to $M_{500} \propto M_{\rm gas}^{0.74 \pm 0.05}$ \cite{Gonzalez} and $M_{500} \propto M_{\rm gas}^{0.88 \pm 0.13}$ \cite{Lagana}. These studies consider the gravitational mass and hot gas mass at a very large radius $r_{500}$, which represent the total gravitational mass and total hot gas mass respectively. Therefore, the correlations are analogous to the Tully-Fisher relation in galaxies, but not the RAR. However, since it is quite hard to determine the actual radius of a galaxy cluster, the resulting uncertainties of the total gravitational mass determination are usually very large and the correlation between the total gravitational mass and total hot gas mass is somewhat weaker and less robust. We use the data in \cite{Chen} and plot $M_{500}$ against $M_{\rm gas}$ for our 64 sample clusters (see Fig.~4). The resulting correlation $M_{500} \propto M_{\rm gas}^{0.73 \pm 0.05}$ is very close to the correlation at intermediate radii. The resulting scatter is 0.15 dex and the correlation coefficient is 0.88. Therefore, the correlation at intermediate radii has a smaller scatter and a larger correlation coefficient than that at $r_{500}$. This is an interesting result because the correlation at smaller radii should be more affected by astrophysical processes like mergers. Therefore, our study focusing on the correlation between the enclosed gravitational mass and hot gas mass at intermediate radii can give a new picture for us to understand the interplay between dark matter and baryons in galaxy clusters.

Another study applies weak lensing to determine the gravitational mass at a fixed radius 1 Mpc for 50 galaxy clusters \cite{Mahdavi}. The relation obtained is $M_{\rm grav} \propto M_{\rm gas}^{0.82\pm 0.15}$. However, considering a fixed radius is a good choice for very large galaxy clusters only. Based on the galaxy cluster sample we considered, 13 of them have $r_{500} \le 1$ Mpc and some of them even have $r_{500} \le 0.7$ Mpc. Therefore, they are the reasons why we consider the enclosed gravitational mass at $r_c$ and $2r_c$ rather than at $r=r_{500}$ or a fixed radius. Generally speaking, our result $M_{\rm grav} \propto M_{\rm gas}^{0.74 \pm 0.02}$ agrees with the previous findings and it gives a much tighter correlation.

We also compare our result with the study in \cite{McGaugh4}, which presents a correlation between baryonic and gravitational masses of cosmic structures spanning a dozen decades in detected baryonic mass (including dwarf galaxies, spiral galaxies and galaxy clusters) (see Fig.~5). The correlation can be best described by a power law $M_{\rm grav} \propto M_{\rm bar}^{0.78 \pm 0.01}$ with a scatter 0.09 dex. By assuming the hot gas mass representing the baryonic mass, we can see that our power-law correlation gives an excellent agreement within the errors with the previous obtained correlation. Since most of the galaxy clusters in our sample are large and massive ($M_{500} \ge 10^{14}M_{\odot}$), the total stellar mass contributes only less than 20\% of the total baryonic mass \citep{Lagana,Gonzalez}. Therefore, the assumption $M_{\rm gas} \approx M_{\rm bar}$ is quite good. It seems that there exists a universal relation between baryonic mass and gravitational masses for different cosmic structures, from dwarf galaxies to galaxy clusters.

\section{Discussion}
Previous studies were not aware of the tight correlation between the enclosed gravitational (or dark matter) mass and hot gas mass in galaxy clusters at intermediate radii ($r \sim 100-1000$ kpc). Some studies investigate the relation between the total gravitational mass and total hot gas mass (or baryonic fraction) of galaxy clusters \cite{Sanders,Vikhlinin2,Mantz,Mahdavi,Gonzalez,Lagana,Donahue,Mantz2}. However, most of the studies are focusing at large radii only. The recent studies about the details of galaxy clusters at $r_{2500}$ did not explicitly discuss the potential correlation \cite{Schellenberger,Eckert}. Also, the rough relations $r_{2500} \sim 0.4r_{500}$ and $r_{500} \sim 5r_c$ give $r_{2500} \sim 2r_c$, which are focusing at relatively larger intermediate radii only. In our analysis, we focus on the enclosed gravitational mass at two different intermediate radii rather than the total gravitational mass. In particular, we need not assume any universal dark matter density profile or cut-off radius to find the gravitational mass. We have used the hydrostatic equation to calculate the mass profile. Although it may include 15-20\% systematic uncertainty, it involves the smallest number of free parameters so that the resultant correlations would be more reliable. Besides, we have followed the single-$\beta$ model, selected larger galaxy clusters and used constant temperature profile to perform the analysis. Generally speaking, these assumptions may potentially lower the scatter of the correlations. Nevertheless, in order to avoid too many parameters involved, we have decided to make these assumptions. Fortunately, the involved uncertainties are not very significant.

The sample clusters considered here may not be the most representative because the analysis originates from a somewhat lower angular resolution data. Also, some systematic bias might be involved because of choosing large galaxy clusters and using the single-$\beta$ model in the analysis. Nevertheless, as mentioned above, the objective of our study is to examine the potential correlation between the hot gas mass and the gravitational mass, but not the accurate quantified relation between them. The systematic deviation between them might slightly affect the scatter in the correlation. On the other hand, recent study shows that the masses of hot gas and stars within dark matter halos of fixed total mass in high mass systems are anti-correlated \cite{Farahi}. This means that it would be better to investigate the correlation between the total mass and the total baryonic mass (hot gas and stellar mass). Further studies in this direction are required to understand the potential interplay between baryonic matter and dark matter.

Recently, the RAR in galaxies reveal a surprising close relation with small scatter between dark matter and baryonic matter from baryon-dominated ($f \approx 1$) to dark-matter-dominated regions ($f \approx 0.09$) \cite{McGaugh3}. In our analysis, although most galaxy clusters do not have baryon-dominated regions, the correlation spans from $f \approx 0.02$ to $f \approx 0.2$, which can be viewed as a close analogy to the RAR. The close connections between dark matter and baryonic matter in galaxies have been discussed for a few decades \cite{Tully,McGaugh,Lelli2,Battaner,Remus,McGaugh2,Lelli,McGaugh3,Li}. As seen in Fig.~5, our power-law correlation also gives an excellent agreement with the correlation between gravitational mass and baryonic mass for various cosmic structures (from dwarf galaxies to galaxy clusters). It is the first time to recognize that such a tight correlation with small scatter also exists in galaxy clusters at intermediate radii and it also agrees with the correlation using the total gravitational and baryonic masses. It provides a new kind of observational evidence to support the universality of correlation between dark matter and baryons. 

Although the gravitational mass and baryonic mass calculated are dependent on the fitted parameters $T$, $r_c$, $\beta$ and $n_0$, these parameters are independent of each other when they are fitted empirically from X-ray observations. One may argue that the parameters $\beta$, $n_0$ and $r_c$ measured in a single fitting procedure of X-ray data might be strongly correlated. However, previous studies obtain $\beta \propto r_c^{0.11^{+0.03}_{-0.02}}$ and $T \propto r_c^{0.03^{+0.05}_{-0.07}}$ \cite{Ota}, which do not show strong and tight correlations among these parameters. We also plot the relation between $n_0$ and $r_c$ in Fig.~6. The correlation is weak and the scatter is 0.27 dex. Since the variation of $\beta$ in our sample clusters is not large ($\beta=0.71 \pm 0.14$), the correlation between $n_0$ and $r_c$ should be strong if there exists any connection between $\beta$, $n_0$ and $r_c$. This means that the intrinsic correlation among these three parameters is not strong. Although some strong relationships such as the $M_{500}-T$ relation \cite{Vikhlinin} and $M_{\rm gas}-T$ relation \cite{Mantz3,Truong} may indicate potential degeneracies among different parameters, these relationships do not necessarily imply there are strong intrinsic correlation among different parameters. For example, the strong relation of $M_{500}$ and $T$ originates from the arbitrary cut-off radii $r_{500}$. Since $M_{500} \propto \beta Tr_{500}$ and the definition of $r_{500}$ gives $r_{500} \propto (\beta T)^{1/2}$, we can easily obtain $M_{500} \propto T^{3/2}$ due to the small range of $\beta$. However, since we are not focusing on the total gravitational mass, but the enclosed gravitational mass at different intermediate radii, the alleged degeneracies of parameters due to this linkage are significantly weakened as no $r_{500}$ is involved. Moreover, the enclosed gravitational mass and baryonic mass at intermediate radii depend on different independent parameters (e.g. $M_{\rm grav}$ depends on $T$ while $M_{\rm gas}$ depends on $n_0$). The strong relation between $M_{500}$ and $T$ does not have any implication on the correlation. It is true that the correlation between $M_{\rm gas}$ and $T$ is very tight ($M_{\rm gas} \propto T^2$ with scatter $=0.10$ dex). However, this relation would give $n_0r_c^3 \propto T^2$ (based on small variation of $\beta$), but not the $M_{\rm grav}-M_{\rm gas}$ correlation. The $M_{\rm grav}-M_{\rm gas}$ correlation would imply something like $T \propto n_0r_c^2$. As shown above, the relation between $T$ and $r_c$ is very weak. If the $M_{\rm grav}-M_{\rm gas}$ correlation is derived from $M_{\rm gas}-T$ relation, then the resulting scatter should be much larger than 0.10 dex. Therefore, our result of low scatter correlation may require some other explanations.

Besides, the potential correlations among the errors of the parameters may also contribute some effects in our analysis. The errors of the parameters ($\beta$, $r_c$ and $n_0$) in the single-$\beta$ model are somewhat correlated. For instance, the percentage error of $\beta$ increases with that of $r_c$ and $n_0$. These potential correlations among the parameters' errors would affect the scatter budget calculated in Table 1. However, these potential correlations might originate from systematic reasons like fittings of the parameters or uncertainties in observations. Since it is difficult to differentiate and quantify the intrinsic correlations among the errors from the observed potential correlations, we did not account for this effect in our calculations. Due to the potential correlations of the parameters' errors, the total expected scatter calculated in Table 1 might be slightly overestimated.

Based on the above arguments, the close connection among these parameters may reveal some kind of interplay between dark matter and baryonic matter during the structure formation. However, based on the CDM model, neither cosmological baryon fraction nor standard gravitation can satisfactorily explain the empirical power-law correlation and its low scatter. The deviation in the dark-matter-dominated regime suggests that baryons might be able to alter the distribution of dark matter. Recent simulations suggest that baryonic processes such as feedbacks from supernovae can alter the dark matter density to form core-like structures \cite{Governato,Pontzen}. However, these processes might be significant in galaxies, but not in galaxy clusters because the amount of baryons is quite insignificant ($f \sim 0.1$) to affect the gravitational potential and the baryonic feedbacks in the intergalactic hot gas medium are nearly negligible. Some numerical simulations show that strong baryonic feedbacks such as AGN feedback may be able to alter the hot gas structure as well as the dark matter distribution \cite{LeBrun,Truong}. However, whether dark matter could be rigorously controlled by the much smaller amount of normal matter is still controversial \cite{Merritt}. Therefore, it is still possible that some `dark sector' physics characterizes the interplay between dark matter and baryonic matter.

In fact, recent studies suggest that dark matter-baryon interaction can explain the excess absorption of the 21-cm signal at cosmic dawn \cite{Bowman,Barkana}. Some studies also suggest that dark matter-baryon interaction can explain the observed MDA relation \cite{Famaey} and the relation between dark matter density and disk length scales \cite{Salucci,Chan2}. It seems that the possible interactions between dark matter and baryons might be able to account for the close connection between gravitational mass and hot gas mass for both galaxies and galaxy clusters. Moreover, some recent studies apply modified gravity or modified Newtonian dynamics (MOND) to explain the one-to-one correspondence between dark matter and baryons in galaxies without the help of dark matter \cite{Milgrom,Lelli3}. The existence of a new dynamical law rather than dark matter may also be a viable way to explain the tight correlation. 

\section{Acknowledgements}
The work described in this paper was supported by a grant from the Research Grants Council of the Hong Kong Special Administrative Region, China (Project No. EdUHK 28300518).




\begin{thebibliography}{00}
\bibitem{Tully} Tully, R. B., Fisher, J. R. A new method of determining distances to galaxies. Astron. Astrophys. 54, 661 (1977).
\bibitem{McGaugh} McGaugh, S. S. The Baryonic Tully-Fisher Relation of Gas-rich Galaxies as a Test of $\Lambda$CDM and MOND. Astron. J. 143, 40 (2012).
\bibitem{Lelli2} Lelli, F., McGaugh, S. S., Schombert, J. M. The small scatter of the baryonic Tully-Fisher relation. Astrophys. J. 816, L14 (2016).
\bibitem{Battaner} Battaner, E., Florido, E. The rotation curve of spiral galaxies and its cosmological implications. Fund. Cosmic. Phys. 21, 1 (2000).
\bibitem{Remus} Remus, R.-S., Burkert, A., Dolag, K., Johansson, P. H., Naab, T., Oser, L., Thomas, J. The dark halo-spheroid conspiracy and the origin of elliptical galaxies. Astrophys. J. 766, 71 (2013).
\bibitem{McGaugh2} McGaugh, S. S. The Mass Discrepancy-Acceleration Relation: Disk Mass and the Dark Matter Distribution. Astrophys. J. 609, 652 (2004).
\bibitem{Lelli} Lelli, F., McGaugh, S. S., Schombert, J. M., Pawlowski M. S. The Relation between Stellar and Dynamical Surface Densities in the Central Regions of Disk Galaxies. Astrophys. J. 827, L19 (2016).
\bibitem{McGaugh3} McGaugh, S. S., Lelli, F., Schombert, J. M. Radial Acceleration Relation in Rotationally Supported Galaxies. Phys. Rev. Lett. 117, 201101 (2016).
\bibitem{Li} Li, P., Lelli, F., McGaugh, S. S., Schombert, J. Fitting the radial acceleration relation to individual SPARC galaxies. Astron. Astrophys. 615, A3 (2018). 
\bibitem{Desmond} Desmond, H. A statistical investigation of the mass discrepancy acceleration relation. Mon. Not. R. Astron. Soc. 464, 4160 (2017).
\bibitem{Chan} Chan, M. H. Analytic expressions for the dark matter-baryon relations. Int. J. Mod. Phys. D 26, 1750118 (2017).
\bibitem{Ludlow} Ludlow, A. D. {\it et al.} Mass-Discrepancy Acceleration Relation: A Natural Outcome of Galaxy Formation in Cold Dark Matter Halos. Phys. Rev. Lett. 118, 161103 (2017).
\bibitem{Springel} Springel, V. The cosmological simulation code GADGET-2. Mon. Not. R. Astron. Soc. 364, 1105 (2005).
\bibitem{Planelles} Planelles, S., Borgani, S., Dolag, K., Ettori, S., Fabjan, D., Murante, G., Tornatore, L. Baryon census in hydrodynamical simulations of galaxy clusters. Mon. Not. R. Astron. Soc. 431, 1487 (2013).
\bibitem{Lin} Lin, Y.-T., Mohr, J. J., Stanford, S. A., Near-Infrared Properties of Galaxy Clusters: Luminosity as a Binding Mass Predictor and the State of Cluster Baryons. Astrophys. J. 591, 749 (2003).
\bibitem{Giodini2} Giodini, S. {\it et al.}. Stellar and Total Baryon Mass Fractions in Groups and Clusters Since Redshift 1. Astrophys. J. 703, 982 (2009).
\bibitem{Lagana2} Lagan\'a, T. F., Zhang, Y.-Y., Reiprich, T. H., Schneider, P. XMM-Newton/Sloan Digital Sky Survey: Star Formation Efficiency in Galaxy Clusters and Constraints on the Matter-density Parameter. Astrophys. J. 743, 13 (2011).
\bibitem{Sanders} Sanders, R. H. The virial discrepancy in clusters of galaxies in the context of modified Newtonian dynamics. Astrophys. J. 512, L23 (1999).
\bibitem{Vikhlinin2} Vikhlinin, A. {\it et al.}. Chandra Cluster Cosmology Project. II. Samples and X-Ray Data Reduction. Astrophys. J. 692, 1033 (2009).
\bibitem{Mantz} Mantz, A., Allen, S. W., Ebeling, H., Rapetti, D., Drlica-Wagner, A. The observed growth of massive galaxy clusters-II. X-ray scaling relation. Mon. Not. R. Astron. Soc. 406, 1773 (2010).
\bibitem{Mahdavi} Mahdavi, A., Hoekstra, H., Babul, A., Bildfell, C., Jeltema, T., Henry, J. P. Joint Analysis of Cluster Observations. II. Chandra/XMM-Newton X-Ray and Weak Lensing Scaling Relations for a Sample of 50 Rich Clusters of Galaxies. Astrophys. J. 767, 116 (2013).
\bibitem{Donahue} Donahue, M. {\it et al.}. CLASH-X: A Comparison of Lensing and X-Ray Techniques for Measuring the Mass Profiles of Galaxy Clusters. Astrophys. J. 794, 136 (2014).
\bibitem{Mantz2} Mantz, A. B., Allen, S. W., Morris, R. G., Schmidt, R. W. Cosmology and astrophysics from relaxed galaxy clusters - III. Thermodynamic profiles and scaling relations. Mon. Not. R. Astron. Soc. 456, 4020 (2016).
\bibitem{Schellenberger} Schellenberger, G., Reiprich, T. H. HICOSMO - cosmology with a complete sample of galaxy clusters. I. Data analysis, sample selection and luminosity-mass scaling-relation. Mon. Not. R. Astron. Soc. 469, 3738 (2017).
\bibitem{Eckert} Eckert, D. {\it et al.} Non-thermal pressure support in X-COP galaxy clusters. Astron. Astrophys. 621, A40 (2019).
\bibitem{Ota} Ota, N., Mitsuda, K. A uniform X-ray analysis of 79 distant galaxy clusters with ROSAT and ASCA. Astron. Astrophys. 428, 757 (2004).
\bibitem{Chen} Chen, Y., Reiprich, T. H., B\"ohringer, H., Ikebe, Y., Zhang, Y.-Y. Statistics of X-ray observables for the cooling-core and non-cooling core galaxy clusters.  Astron. Astrophys. 466, 805 (2007).
\bibitem{Sun} Sun, M., Voit, G. M., Donahue, M., Jones, C., Forman, W., Vikhlinin, A. Chandra Studies of the X-Ray Gas Properties of Galaxy Groups. Astrophys. J. 693, 1142 (2009).
\bibitem{Ventimiglia} Ventimiglia, D. A., Voit, G. M., Rasia, E. Temperature Structure and Mass-Temperature Scatter in Galaxy Clusters. Astrophys. J. 747, 123 (2012).
\bibitem{Lagana} Lagan\'a, T. F., Martinet, N., Durret, F., Neto, G. B. L., Maughan, B., Zhang, Y.-Y. A comprehensive picture of baryons in groups and clusters of galaxies. Astron. Astrophys. 555, A66 (2013).
\bibitem{Giodini} Giodini, S., Lovisari, L., Pointecouteau, E., Ettori, S., Reiprich, T. H., Hoekstra, H. Scaling Relations for Galaxy Clusters: Properties and Evolution. Sp. Sci. Rev. 177, 247 (2013).
\bibitem{Ettori} Ettori, S. The generalized scaling relations for X-ray galaxy clusters: the most powerful mass proxy. Mon. Not. R. Astron. Soc. 435, 1265 (2013).
\bibitem{Reiprich} Reiprich, T. H., B\"ohringer, H. The Mass Function of an X-Ray Flux-limited Sample of Galaxy Clusters. Astrophys. J. 567, 716 (2002).
\bibitem{Hoekstra} Hoekstra, H., Herbonnet, R., Muzzin, A., Babul, A., Mahdavi, A., Viola, M., Cacciato, M. The Canadian Cluster Comparison Project: detailed study of systematics and updated weak lensing masses. Mon. Not. R. Astron. Soc. 449, 685 (2015).
\bibitem{Biffi} Biffi, V. {\it et al.} On the Nature of Hydrostatic Equilibrium in Galaxy Clusters. Astrophys. J. 827, 112 (2016).
\bibitem{Bartalucci} Bartalucci, I., Arnaud, M., Pratt, G. W., Le Brun, A. M. C. Possible interaction between baryons and dark-matter particles revealed by the first stars. Astron. Astrophys. 617, A64 (2018).
\bibitem{Churazov} Churazov, E., Forman, W., Jones, C., B\"ohringer, H. XMM-Newton Observations of the Perseus Cluster. I. The Temperature and Surface Brightness Structure. Astrophys. J. 590, 225 (2003).
\bibitem{Sanderson} Sanderson, A. J. R., Ponman, T. J., O'Sullivan, E. A statistically selected Chandra sample of 20 galaxy clusters-I. Temperature and cooling time profiles. Mon. Not. R. Astron. Soc. 372, 1496 (2006).
\bibitem{Hudson} Hudson, D. S., Mittal, R., Reiprich, T. H., Nulsen, P. E. J., Andernach H., Sarazin C. L. What is a cool-core cluster? A detailed analysis of the cores of the X-ray flux-limited HIFLUGCS cluster sample. Astron. Astrophys. 513, A37 (2010).
\bibitem{Vikhlinin} Vikhlinin, A., Kravtsov, A., Forman, W., Jones, C., Markevitch, M., Murray, S. S., Van, Speybroeck L. Chandra Sample of Nearby Relaxed Galaxy Clusters: Mass, Gas Fraction, and Mass-Temperature Relation. Astrophys. J. 640, 691 (2006).
\bibitem{Reiprich2} Reiprich, T. H., Basu, K., Ettori, S., Israel, H., Lovisari, L., Molendi, S., Pointecouteau, E., Roncarelli, M. Outskirts of Galaxy Clusters. Sp. Sci. Rev. 177, 195 (2013).
\bibitem{Ikebe} Ikebe, Y. X-ray Statistical Properties of the Central Cool Component in Clusters of Galaxies. {\it Proceedings of the MPA/ESO/MPE/USM Joint Astronomy Conference}, 81 (astro-ph/0112132) (2002).
\bibitem{Rossetti} Rossetti, M., Gastaldello, F., Eckert, D., Della Torre, M., Pantiri, G., Cazzoletti, P., Molendi, S. The cool core state of Planck SZ-selected clusters versus X-ray selected samples: evidence for cool core bias. Mon. Not. R. Astron. Soc. 468, 1917 (2017).
\bibitem{Teerikorpi} Teerikorpi, P. Eddington-Malmquist bias in a cosmological context. Astron. Astrophys. 576, A75 (2015).
\bibitem{Kafer} K\"afer, F., Finoguenov, A., Eckert, D., Sanders, J. S., Reiprich, T. H., Nandra, K. Towards a characterization of X-ray galaxy clusters for cosmology. ArXiv:1907.03806. 
\bibitem{Planck} Ade, P. A. R. {\it et al.}. Planck 2013 results. I. Overview of products and scientific results. Astron. Astrophys. 571, A1 (2014).
\bibitem{Cavaliere} Cavaliere, A., Lapi, A. The Astrophysics of the Intracluster Plasma. Phys. Rept. 533, 69 (2013).
\bibitem{Navarro} Navarro, J. F., Frenk, C. S., White, S. D. M. A Universal Density Profile from Hierarchical Clustering. Astrophys. J. 490, 493 (1997).
\bibitem{Javid} Javid, K., Perrott, Y. C., Rumsey, C., Saunders, R. D. E. Physical Modelling of Galaxy Clusters Using Einasto Dark Matter Profiles. Mon. Not. R. Astron. Soc. 489, 3135 (2019).
\bibitem{Gonzalez} Gonzalez, A. H., Sivanandam, S., Zabludoff, A. I., Zaritsky, D. Galaxy Cluster Baryon Fractions Revisited. Astrophys. J. 778, 14 (2013).
\bibitem{McGaugh4} McGaugh, S. S., Schombert, J. M., de Blok, W. J. G. \& Zagursky, M. J. The Baryon Content of Cosmic Structures. Astrophys. J. 708, L14 (2010).
\bibitem{Farahi} Farahi, A. {\it et al.} Detection of anti-correlation of hot and cold baryons in galaxy clusters. Nature Comm. 10, 2504 (2019).
\bibitem{Mantz3} Mantz, A. B. {\it et al.} Weighing the giants V: Galaxy cluster scaling relations. Mon. Not. R. Astron. Soc. 463, 3582 (2016).
\bibitem{Truong} Truong, N. {\it et al.} Cosmological hydrodynamical simulations of galaxy clusters: X-ray scaling relations and their evolution. Mon. Not. R. Astron. Soc. 474, 4089 (2018).
\bibitem{Governato} Governato, F. {\it et al.} Bulgeless dwarf galaxies and dark matter cores from supernova-driven outflows. Nature 463, 203 (2010).
\bibitem{Pontzen} Pontzen, A., Governato, F. Cold dark matter heats up. Nature 506, 171 (2014).
\bibitem{LeBrun} Le Brun, A. M. C., McCarthy, I. G., Schaye, J., Ponman, T. J. The scatter and evolution of the global hot gas properties of simulated galaxy cluster populations. Mon. Not. R. Astron. Soc. 466, 4442 (2017).
\bibitem{Merritt} Merritt, D. Cosmology and convention. Studies in History and Philosophy of Modern Physics 57, 41 (2017).
\bibitem{Bowman} Bowman, J. D., Rogers, A. E. E., Monsalve, R. A., Mozdzen, T. J., Mahesh, N. An absorption profile centred at 78 megahertz in the sky-averaged spectrum. Nature 555, 67 (2018).
\bibitem{Barkana} Barkana, R. Possible interaction between baryons and dark-matter particles revealed by the first stars. Nature 555, 71 (2018).
\bibitem{Famaey} Famaey, B., Khoury, J., Penco, R. Emergence of the mass discrepancy-acceleration relation from dark matter-baryon interactions. J. Cosmol. Astropart. Phys. 03, 038 (2018).
\bibitem{Salucci} Salucci, P., Turini, N. Evidences for Collisional Dark Matter In Galaxies? arXiv:1707.01059.
\bibitem{Chan2} Chan, M. H. A universal constant for dark matter-baryon interplay. Sci. Rept. 9, 3570 (2019).
\bibitem{Milgrom} Milgrom, M. Universal Modified Newtonian Dynamics Relation between the Baryonic and Dynamical Central Surface Densities of Disc Galaxies. Phys. Rev. Lett. 117, 141101 (2016).
\bibitem{Lelli3} Lelli, F., McGaugh, S. S., Schombert, J. M. Testing Verlinde's Emergent Gravity with the Radial Acceleration Relation. Mon. Not. R. Astron. Soc. 468, L68 (2017).
\end{thebibliography}


\end{document}